\documentclass[twocolumn]{aastex63}
\usepackage{lineno}
\usepackage{amsmath}
\usepackage{amssymb}
\usepackage{epsfig}
\usepackage{natbib}
\usepackage{graphicx}
\usepackage[figuresright]{rotating}
\usepackage{color}
\usepackage{multirow}
\usepackage{booktabs}
\usepackage{subfigure}
\usepackage{ulem}
\usepackage{url}
\def\UrlBreaks{\do\A\do\B\do\C\do\D\do\E\do\F\do\G\do\H\do\I\do\J
\do\K\do\L\do\M\do\N\do\O\do\P\do\Q\do\R\do\S\do\T\do\U\do\V
\do\W\do\X\do\Y\do\Z\do\[\do\\\do\]\do\^\do\_\do\`\do\a\do\b
\do\c\do\d\do\e\do\f\do\g\do\h\do\i\do\j\do\k\do\l\do\m\do\n
\do\o\do\p\do\q\do\r\do\s\do\t\do\u\do\v\do\w\do\x\do\y\do\z
\do\.\do\@\do\\\do\/\do\!\do\_\do\|\do\;\do\>\do\]\do\)\do\,
\do\?\do\'\do+\do\=\do\#}

\newcommand       \G            {{\rm G}}

\newcommand       \mum        {\,{\rm \mu m}}
\newcommand       \Ks           {{ K_{\rm S}}}
\newcommand       \K             {\,{\rm K}}

\newcommand{\AV}{A_{\rm V}}
\newcommand{\Av}{A_{\rm V}}

\newcommand{\AKs}{A_{\rm K_S}}

\newcommand{\AJ}{A_{\rm J}}

\newcommand{\Rv}{R_{\rm V}}

\shorttitle{JWST $0.6-5.3\mum$ Extinction Law}
\shortauthors{Wang \& Chen}

\begin{document}

\title{The Optical to Infrared $0.6-5.3\mum$ Dust Extinction Law of the Milky Way with JWST NIRSpec: Westerlund 2}

\correspondingauthor{Shu Wang}
\email{shuwang@nao.cas.cn}

\author[0000-0003-4489-9794]{Shu Wang}
\affiliation{CAS Key Laboratory of Optical Astronomy, National Astronomical Observatories, 
Chinese Academy of Sciences, Beijing 100101, China}

\author[0000-0001-7084-0484]{Xiaodian Chen}
\affiliation{CAS Key Laboratory of Optical Astronomy, National Astronomical Observatories, 
Chinese Academy of Sciences, Beijing 100101, China}
\affiliation{Institute for Frontiers in Astronomy and Astrophysics, Beijing Normal University,  Beijing 102206, China} 
\affiliation{School of Astronomy and Space Science, University of the Chinese Academy of Sciences, Beijing 101408, China}

\begin{abstract} 
The interstellar extinction law is important for interpreting observations and inferring the properties of interstellar dust grains. 
Based on the 993 prism/CLEAR spectra from the James Webb Space Telescope (JWST), we investigate $0.6-5.3\mum$ interstellar dust extinction law. We propose a pair method to obtain the reddening curves based only on JWST observed spectra. Most of the high extinction sources are toward the young star cluster Westerlund 2. 
The infrared (IR) $1.0-5.3\mum$ reddening curves agree with the power-law $A_\lambda \propto \lambda^{-\alpha}$ well.
We determine an average value of $\alpha=1.98\pm0.15$, which is consistent with the average value of the Galaxy. We find that $\alpha$ may be variable and independent of $\Rv$. 
With the derived $\alpha$, we convert the reddening curves into the extinction curves and establish the non-parameterized $\alpha$-dependent extinction curves in the wavelength range of $0.6-5.3\mum$. At $\lambda<1\mum$, the derived extinction law is not well described by the parameterized power-law type curve. Our non-parameterized $\alpha$-dependent extinction curves are suitable for the extinction correction of JWST-based photometry and spectra measurements at $0.6-5.3\mum$.
We also provide the extinction coefficients for the JWST NIRCam bandpasses with different $\alpha$.   
\end{abstract}

\keywords{Interstellar dust extinction (837); Interstellar reddening (853); Reddening law (1377); Interstellar extinction (841); Star clusters (1567);}

\section{Introduction}\label{intro}

With the increasing wealth of optical spectroscopic, astrometric, and photometric data, the optical extinction law has been relatively well understood in recent years \citep{2016ApJ...821...78S, 2019ApJ...886..108F, 2019ApJ...877..116W, 2023ApJ...950...86G, 2023ApJ...956...26L, 2023ApJ...946...43W, 2023ApJS..269....6Z}. 
However, the infrared (IR) extinction law is not well understood and is still controversial. 
In the last century, it has been reported that the near-IR extinction curve of 0.9 to $3\mum$ can be approximately characterized by a universal power-law $A_\lambda \propto \lambda^{-\alpha}$ with $\alpha \approx 1.6-1.8$ \citep{1985ApJ...288..618R, 1989ApJ...345..245C, 1989ESASP.290...93D, 1990ApJ...357..113M, 1993ApJ...408..573W, 1995ApJS..101..335H}. 
However, a large number of steeper near-IR extinction curves with $\alpha>1.9$ or even $\sim2.6$ have been reported in different lines of sight in the last two decades \citep{2005A&A...435..575M, 2006ApJ...638..839N, 2009ApJ...696.1407N, 2009MNRAS.394.2247G, 2009MNRAS.400..731S, 2010A&A...511A..18S, 2011ApJ...737...73F, 2014ApJ...788L..12W, 2017ApJ...849L..13A, 2018ApJ...859..137C, 2019A&A...630L...3N, 2020MNRAS.496.4951M}.  
In addition, several studies argued that the power-law index $\alpha$ depends on the wavelength region \citep[e.g.][]{2010A&A...511A..18S, 2019A&A...630L...3N}. 
Based on these different $\alpha$ values, the resultant near-IR relative extinction $\AJ/\AKs$ values are from 2.5 to 3.5 with an uncertainty of at least 20\%.

Whether near-IR extinction curve varies with sightlines or the optical extinction is also controversial. 
Some studies support no significant variation \citep{2005ApJ...619..931I, 2009MNRAS.400..731S, 2014ApJ...788L..12W, 2015ApJ...811...38W, 2019A&A...630L...3N, 2020MNRAS.496.4951M}. 
In contrast, some studies found that the near-IR extinction curve changes from one sightline to another \citep{2006ApJ...638..839N, 2009MNRAS.394.2247G, 2017ApJ...849L..13A, 2022ApJ...930...15D}.  
The IR extinction laws are important for constraining dust properties \citep{2015ApJ...811...38W, 2015MNRAS.454..569W, 2021ApJ...906...73H}. 

Near-IR extinction studies are mainly based on broadband photometric data from ground-based telescopes, such as Two Micron All-Sky Survey (2MASS), United Kingdom Infrared Deep Sky Survey (UKIDSS), VISTA Variables in the V\'ia L\'actea survey (VVV). 
Recently, \citet{2023A&A...671L..14F} measured the near-IR extinction law of the LMC 30 Doradus based on photometric data from the JWST Near-Infrared Camera (NIRCam).  
Very few works have been done based on spectroscopic observations. \citet{2011ApJ...737...73F} derived the Galactic center 1 to $19\mum$ extinction curve, where the near-IR $1.2-2.4\mum$ data are from the SINFONI imaging spectroscopy. 
\citet{2022ApJ...930...15D} measured the $0.8-5.5\mum$ extinction curves of 15 O-, B-type star lines of sight by using IRTF/SpeX spectra.

The improvement of the accuracy of extinction laws requires data with higher precision and depth.
There is still great potential for improving the accuracy of IR extinction curves.
The Near Infrared Spectrograph (NIRSpec) on the JWST provides multi-target and integral field spectroscopy at different resolutions ($\sim$100, $\sim$1,000, and $\sim$2,700) over a wide wavelength range from 0.6 to $5.3\mum$.
The prism offers a low-resolution data ($\sim$100) with full wavelength coverage.  
This provides an opportunity to investigate the interstellar extinction law. 
In this work, we use the available NIRSpec prism spectroscopic data to explore the optical to IR $0.6-5.3\mum$ dust extinction law. 
We describe the JWST data and their basic reduction in Section~\ref{data}. 
In Section~\ref{method}, we present the method and results of measuring the reddening curves and the extinction curves. 
The discussion and comparison are in Section~\ref{discussion}. 
The estimated extinction coefficients for bandpasses of the JWST NIRcam are also presented in Section~\ref{discussion}. 
We summarize our main results in Section~\ref{conclusion}.

\section{Data and Sample} \label{data}
\subsection{JWST Data} 

We used publicly available JWST NIRSpec prism/CLEAR spectra. The spectra of the prism disperser and CLEAR filter cover the full wavelengths from 0.6 to $5.3\mum$ with a resolution of $\sim100$. 
In the following, we briefly describe how to obtain and process these spectra. 

Firstly, we searched all the available prism/CLEAR spectra for stars in the JWST observations. 
In total, we obtained 993 spectra
\footnote{The data used in this work can be found in MAST: \dataset[https://doi.org/10.17909/0f0s-0531]{https://doi.org/10.17909/0f0s-0531}.}. 
These spectra are belonged to four programs, including IDs: 1448 (PI: E. Smith), 1473 (PI: M. Perrin), 1493 (PI:  C. R. Proffitt), and 2640 (PI: W. Best). 
Then, we downloaded the level 3 data products from the Barbara A. Mikulski Archive for Space Telescopes (MAST)\footnote{https://mast.stsci.edu} Portal Download Manager. 
The level 3 data products are combined, calibrated science products, and we used them for the next step in the analysis, given that we plan to study extinction laws from a large sample. Since the wavelengths of the different spectra differ slightly, we interpolated the spectra using a uniform wavelength range from 0.6 to $5.3\mum$ in a step of $0.001\mum$.
After that, we examined the fluxes of these spectra and eliminated the sources with negative fluxes. Finally, 627 spectra remained, most of them are M dwarfs or brown dwarfs. For dwarfs with higher extinction, they are in the Westerlund 2 region. While for dwarfs with lower extinction, they are field stars in the Milky Way.

Westerlund 2 \citep{1961ArA.....2..419W} is one of the most massive young star clusters in the Milky Way. 
It is located in the Carina$-$Sagittarius spiral arm with ($l, b$)=($284^{\circ}.3, -0^{\circ}.34$) and is surrounded by molecular clouds. 
The reported distance of Westerlund 2 ranges from 2.8 kpc to 8 kpc \citep[e.g.,][and references therein]{2004MNRAS.347..625C, 2011A&A...535A..40R, 2013A&A...555A..50C, 2015AJ....150...78Z}. 

Westerlund 2 is in a cloudy environment, the cluster members are heavily and differentially reddened with $1.5<E(B-V)<2.1$ mag \citep{2004MNRAS.347..625C, 2015AJ....150...78Z}. 
The average extinction is $\AV=6.51$ mag \citep{2013AJ....145..125V}. 
Based on the pair method or the zero-age main sequence fitting method, the total-to-selective extinction ratio $\Rv$ is derived, such as $\Rv=3.77$ \citep{2013AJ....145..125V}, 3.85 \citep{2013A&A...555A..50C}, 3.95 \citep{2015AJ....150...78Z}, 3.96 \citep{2015MNRAS.450.3855M}, 4.14 \citep{2015MNRAS.446.3797H}, and 3.5-4.5 \citep{2015MNRAS.450.3855M}. 
Considering the large extinction as well as the uncertainty in the extinction law, Westerlund 2 is a good target for studying the optical to IR extinction law.

\subsection{The Sample}\label{sample}
The pair method compares the spectral energy distributions (SEDs) of the same type stars with and without extinction. Usually, spectra without extinction are built from theoretical models based on stellar atmospheric parameters. Considering the difficulty of obtaining atmospheric parameters and the uncertainties in the IR spectra of red stars, in this work, we try to build them based on observed JWST spectra with very small or even negligible extinction. In this section, we construct a sample of extinction pairs that have large extinction differences and match well after extinction correction.

To identify extinction pairs, the extinction information is needed. Therefore, we first estimate the $V$-band extinction difference $\Delta\Av$ of each pair by matching SEDs. 
The differences in stellar SEDs arise from the extinction, as well as the distance and physical radius of the star. 
To eliminate the effect of radius and distance differences, we multiply each original SED $F_\lambda^\star$ by a factor that equalizes the flux at $5\mum$ for all SEDs, and obtain the processed SEDs $F'_\lambda$. The idea is that the extinction at $5\mum$ is smaller than the extinction at short wavelengths. 
After this step, the remaining differences in SEDs come mainly from the extinction. 
With these processed SEDs $F'_\lambda$, we estimate the extinction difference $\Delta\Av$ to select candidate extinction pairs. 
We assume that $\Delta\Av$ ranges from 0 to 50 mag and adopted a priori $\Rv=3.1$ extinction law of \citet{2019ApJ...877..116W} to obtain the extinction covering the complete wavelength range from 0.6 to $5.3\mum$. 

We matched all 627 processed SEDs $F'_\lambda$ one by one and obtain a $627\times627$ matrix.  
For each extinction pair, we calculated the minimum root mean square error (RMSE) by ${\rm RMSE}= \log \sqrt{\sum{(1- F'_{i,n}/F'_{j,n}\times10^{-0.4\Delta{A_n}})^2}}$ and the corresponding extinction $\Delta\Av$. Here, $i$ and $j$ are the indexes of an extinction pair, $n$ is the index of the wavelength, and $\Delta{A_n}$ is estimated by multiplying $\Delta\Av$ by the extinction law. In the matching, we only used fluxes from $1\mum$ to $4\mum$. It is because, at shorter wavelengths, the fluxes are lower, while at longer wavelengths, we forced the fluxes to be equal at $5\mum$. Note that when the extinction is large, the $5\mum$ extinction can still be greater than 0.1 magnitude. For this reason, we estimated a first-order correction based on the initial extinction difference $\Delta\Av$  and adjusted the $5\mum$ flux, then performed the matching again to get the updated RMSE and $\Delta\Av$.

\begin{figure*}[ht]
\centering
\includegraphics[angle=0,width=6.5in]{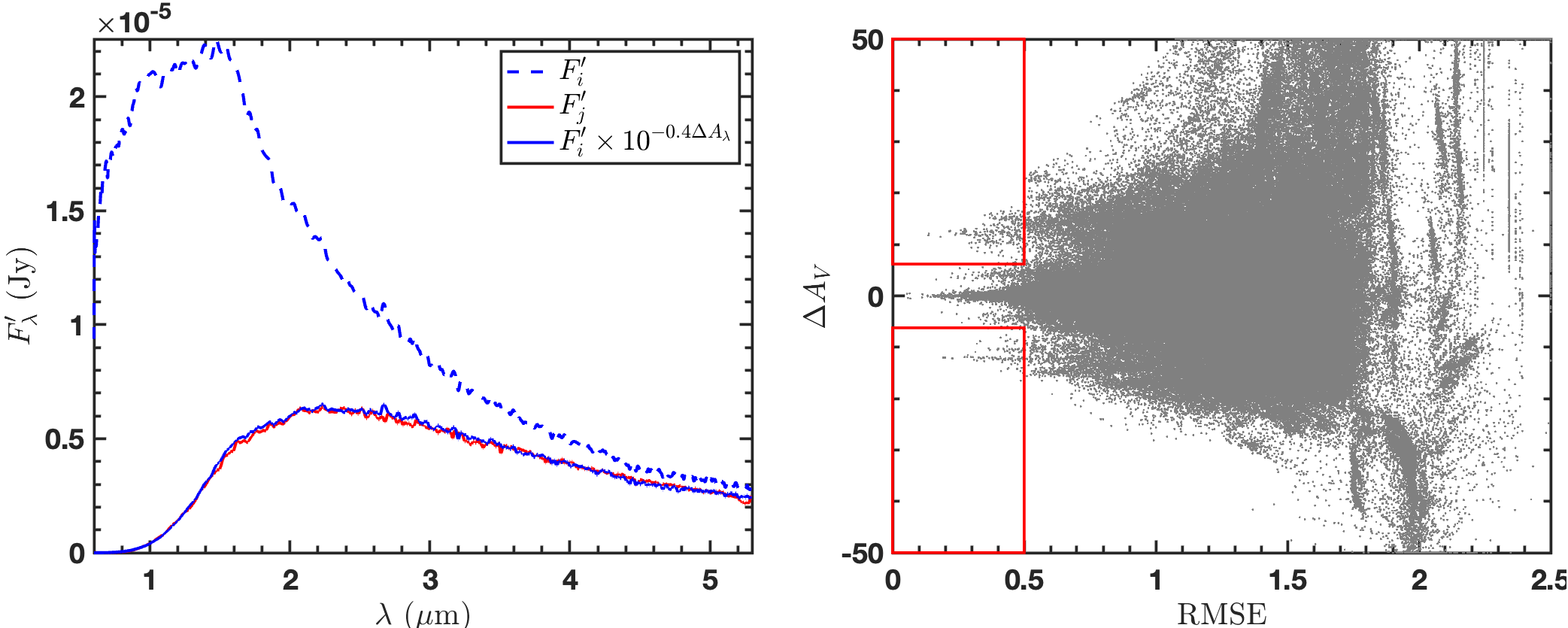}
\caption{
               \label{fig:pair}
The example of selecting extinction pairs. 
Left: an example of SED pairing. $F'_i$ (blue dashed line) and $F'_j$ (red line) are the processed SED of index $i$, $j$, respectively. 
The blue solid line is the extincted SED $F'_i\times 10^{-0.4A_\lambda}$ of index $i$, which matches with the $F'_j$. 
Right: the distribution of $\Delta\AV$ with the RMSE for $627\times627$ pairs. 
The pairs located in the red box area with $|\Delta\Av|>6.2$ mag and RMSE $<0.5$ are selected as the sample for further measuring the extinction law.  
} 
\end{figure*}

The left panel of Figure~\ref{fig:pair} is an example. The blue dashed line is the processed SED $F'_i$ of index $i$. The blue solid line is the extincted SED $F'_i\times 10^{-0.4\Delta{A_n}}$ of index $i$, which matches with the processed SED $F'_j$ of index $j$ (red line). We recorded the corresponding RMSE and $\Delta\AV$ values, which are 0.18 and 11.5 mag for this pair.
After this step, we obtained $627\times627$ pairs. For each pair, we derived the RMSE and $\Delta\AV$ shown in the right panel of Figure~\ref{fig:pair}.  
The candidate extinction pairs need to be well matched and highly extinct. 
Therefore, we selected pairs with $|\Delta\Av|>6.2$ mag and a matching error RMSE $<0.5$, which are located in the red box area in the right panel of Figure~\ref{fig:pair}. The main reason we adopt these criteria is that in the RMSE vs. $\Delta\Av$ plot, the distribution of pairs satisfying these criteria is sharply distinguished from the majority of pairs. 
Finally, we constructed a sample including 408 extinction pairs from the initial $627\times627$ extinction pairs \footnote{ We also tried to select extinction pairs using different selection criteria, e.g., RMSE$<0.3$ with $|\Delta\Av|>6$ mag, RMSE$<0.6$ with $|\Delta\Av|>6$ mag, and $|\Delta\Av|>3$ mag with RMSE$<0.5$, and analyzed the effects on the determination of extinction curves. We found that the average $\alpha$ determined is very consistent ($1.97-2.01$) when different numbers of extinction pairs are selected using different criteria.}. 
Their original SEDs $F_\lambda^\star$ and the corresponding extinction-free SEDs $F_{\lambda}^0$ are obtained as well. 

Note that the processed SEDs $F'_\lambda$ and a priori $\alpha=2.07$ extinction law are only used to estimate $\Delta\Av$ for the selection of candidate extinction pairs and are not used in subsequent extinction curve measurements. The reasons to choose the $\alpha=2.07$ extinction law are discussed in Section~\ref{prioriextlaw}.
The extinction of the sample sources is comparable to the average extinction of the Westerlund 2 region ($\Av\sim7$ mag), and the RMSE $<0.5$ corresponds to an average matching error in flux of less than 5.8\%.

\section{Measuring the Reddening Curve and Extinction Law} \label{method}

In this section, we describe the method and procedure of determining the reddening curve and extinction law. 
First, we measure the reddening curves by matching the SEDs of the extinction pairs. 
After that, we fit the near-IR reddening curves with a power-law to obtain the index $\alpha$. 
We also fit the optical reddening curves with the $\Rv$-parameterized curves to derive $\Rv$ values.  
Finally, we convert the reddening curves into the extinction curves.

\subsection{The Reddening Curve}\label{reddening}

Since the distance information of stars in our sample is not available, we first estimate the color excess ratios (CERs) 
\begin{eqnarray} \label{equ_k}
k_\lambda  = \frac{E(\lambda- {\rm J200})}{E({\rm J115}-{\rm J200})} 
 =   \frac{A_\lambda-A_{\rm J200}}{A_{\rm J115}-A_{\rm J200}}~ 
\end{eqnarray}
to determine the reddening curve. 
This is analogous to the commonly used CER $k_\lambda=E(\lambda-\Ks)/E(J-\Ks)$, but with the monochromatic near-IR 
J115 ($1.15\mum$) and J200 ($2.0\mum$) wavelengths 
substituting for the $J$ and $\Ks$ bands.
This method is similar to that used by \citet{2019ApJ...886..108F} to analyze the ultraviolet to optical extinction curves.

To determine $k_\lambda$ for a given pair, we first estimated $A_\lambda$, $A_{\rm J115}$, and $A_{\rm J200}$.
For each pair, the extinction can be estimated from the original spectra by the equation $A_\lambda=-2.5\log \frac{F_{\lambda,i}^\star R_i^2 d_j^2}{F_{\lambda, j}^\star R_j^2 d_i^2}$. Then, we determined the extinction difference of different wavelengths to eliminate radius $R$ and distance $d$ by $A_\lambda-A_{\rm J200}=-2.5\log \frac{F_{\lambda,i}^\star }{F_{\lambda, j}^\star}+2.5\log \frac{F_{{\rm J200},i}^\star }{F_{{\rm J200}, j}^\star}$. $A_{\rm J115} - A_{\rm J200}$ can be determined similarly. With Equation~\ref{equ_k} and the definition of extinction, we derived CER $k_\lambda$ for 408 pairs. The variation of $k_\lambda$ with $\lambda$ is the reddening curve.

\subsection{The Near-IR Power-Law Index $\alpha$}\label{alpha}

With the obtained 408 reddening curves, we try to convert them into extinction curves. 
Considering that the near-IR extinction curve approximates a power-law (see Section~\ref{intro}), we attempt to fit the near-IR reddening curve with a power-law  $A_\lambda \propto \lambda^{-\alpha}$ and obtain the index $\alpha$. 
Then, the near-IR relative extinction can be derived, such as  
\begin{equation} \label{equ_NIR1}
\frac{A_{\rm J115}}{A_{\rm J200}}=(\frac{\lambda_{\rm J115}}{\lambda_{\rm J200}})^{-\alpha}~~, \\
\end{equation}
\begin{equation} \label{equ_NIR2}
\frac{A_{\rm J200}}{E(\rm J115-J200)}=\frac{\lambda_{\rm J200}^{-\alpha}}{\lambda_{\rm J115}^{-\alpha}-\lambda_{\rm J200}^{-\alpha}}~~. \\
\end{equation} 
Combined the near-IR relative extinction with the CER $k_\lambda$ derived in Section~\ref{reddening}, the extinction coefficients are obtained by
\begin{equation} \label{equ_A/E}
\frac{A_\lambda}{E(\rm J115-J200)} 
 = k_\lambda +  \frac{A_{\rm J200}}{E(\rm J115-J200)}~.  
\end{equation}
Finally, the wavelength-dependent extinction curve is derived.

\begin{figure*}[ht]
\centering
\includegraphics[angle=0,width=3.2in]{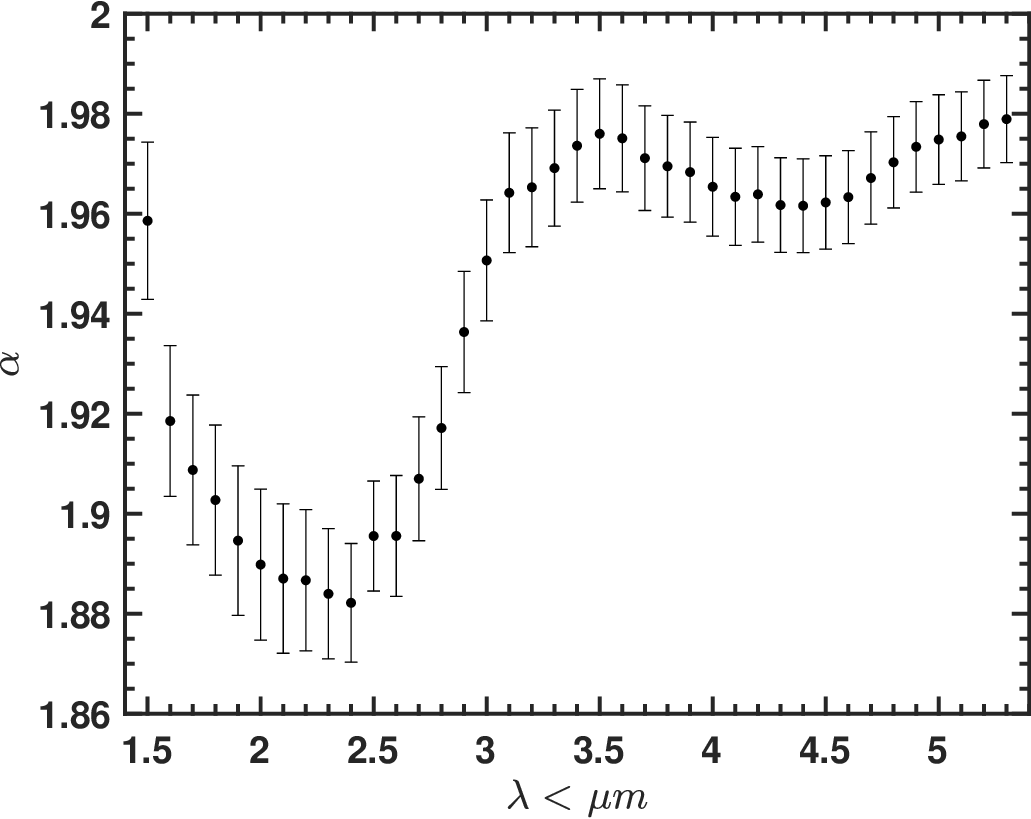}
\hspace{0.5cm}
\includegraphics[angle=0,width=3.2in]{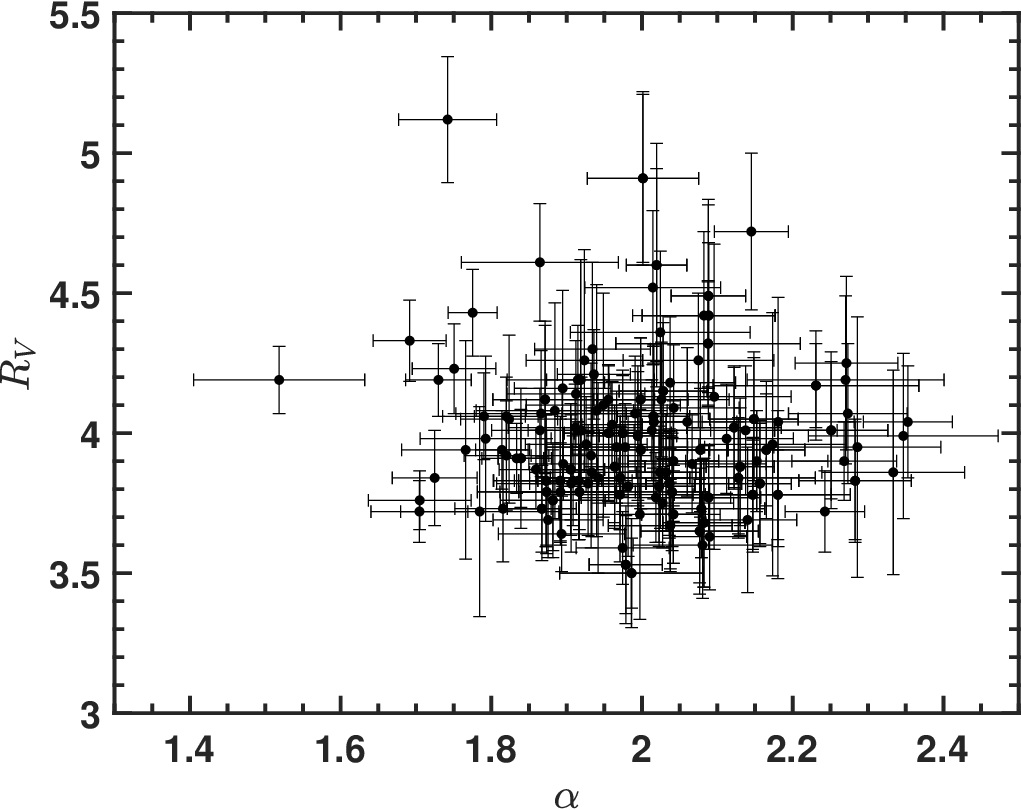}
\caption{
               \label{fig:alpha_lambda}
Left: Distribution of the average near-IR power-law index $\alpha$ with the cutoff wavelength $\lambda$.   
Right: Distribution of the total-to-selective extinction ratio $\Rv$ with the near-IR power-law index $\alpha$.   
} 
\end{figure*}

Although the near-IR extinction curve satisfies the power-law well, in practice, how to choose the near-IR wavelength range of the fit affects the determination of power-law index $\alpha$. 
The extinction curve deviates significantly from the power-law at wavelengths $\lambda<1\mum$, so we set the starting wavelength to $1\mum$ in our fitting. 
For the cutoff wavelength, we used a varying wavelength, considering a range of 1.5 to $5.3\mum$, with a step of 0.1. Each reddening curve is then fitted with a power-law $A_\lambda \propto \lambda^{-\alpha}$ to obtain a set of $\alpha$ values. The left panel of Figure~\ref{fig:alpha_lambda} shows the variation of the average $\alpha$ with cutoff wavelength $\lambda$.  
Generally, the average $\alpha$ values obtained with different cutoff wavelengths range from 1.88 to 1.98, with some systematic effects. With the increasing of the cutoff wavelength, the $\alpha$ value decreases first, then increases, and finally remains unchanged. When the cutoff wavelength is set to $2\mum$, the average $\alpha$ value is 1.89. This may be due to the existence of some details of the extinction law near $2\mum$ that deviate from the power-law (see the bottom panels of Figure~\ref{fig:ext}). 
The statistical error of $\alpha$ decreases as the cutoff wavelength increase, so we decided to fit the $1-5.3\mum$ reddening curve to obtain the index $\alpha$. It should be noted that the choice of different cutoff wavelengths systematically affects the value of $\alpha$ slightly.

The reddening curves used to determine the $\alpha$ values were obtained by using J115 ($1.15\mum$) and J200 ($2.0\mum$) as reference bands (Section~\ref{reddening}). 
In fact, any bands can be used as the reference bands, and the change of reference bands will affect the determination of the $\alpha$ value. 
Therefore, we also discuss the effect of using different reference bands on the measurement of $\alpha$ values.

We fixed one of the reference bands to J115 and changed the other from 1.65 to $5.1\mum$ in a step of $0.05\mum$. 
For each extinction pair, we calculated the $\alpha$ values separately based on the reddening curves obtained by using different reference bands. 
For each extinction pair, we derived 70 $\alpha$ values, the corresponding mean $<\alpha>$ value, and the standard deviation $\sigma_1$. 
For a total of 408 extinction pairs, there are $408\times70$ $\alpha$ values. 
The mean value of the standard deviations $<\sigma_1>$ is 0.09 larger than the statistical error $\sigma_2\sim0.03$ of the power-law fit. 
Finally, we decided to take J200 as the other reference band for two reasons. 
One is that the $\alpha$ value obtained from the reddening curves calculated based on J115 and J200 is close to the mean value of 70 $\alpha$ values. 
The other is that J200 is analogous to the F200W band, a commonly used band for JWST near-IR photometry.

We calculated the CER (Equation~\ref{equ_k}) and obtained the reddening curves based on J115 and J200 in Section~\ref{reddening}. 
In this section, we fit the reddening curves from 1 to $5.3\mum$ with a power-law $A_\lambda \propto \lambda^{-\alpha}$ and obtain the power-law index $\alpha$. 
The $\alpha$ error is estimated by $\sigma_\alpha=\sqrt{\sigma_1^2+\sigma_2^2}$. We selected extinction pairs with $\sigma_\alpha<0.15$, leaving 384 pairs for subsequent analysis. Except for 4 pairs, the high extinction sources in the other pairs are located in the Westerlund 2 region.
The mean $\sigma_\alpha$ of 384 pairs is 0.08. 
The mean $\alpha$ value of 384 pairs is 1.98 with a standard deviation of 0.15. Note that a total of 11 highly reddened stars exist in multiple pairs. Their average $\alpha$ and $\Rv$ values are consistent with those of the whole sample and have slightly smaller dispersion. For example, the dispersion of $\alpha$ is between 0.06 and 0.15.
 
In addition, we examined the correlation between $\alpha$ and the absolute extinction $\Av$. The correlation coefficient between $\alpha$ and $\Av$ is $\rho=-0.0859$ and the associated probability $p=0.0929$, indicating the null hypothesis of significant correlation was rejected. It means that $\alpha$ is independent of the absolute extinction.

\subsection{The $\Rv$ Value}\label{Rv}

We estimate the $\Rv$ value of each extinction pair by matching the reddening curves from $0.61$ to $1\mum$ with the $\Rv$-parameterized extinction curves of \citet{2023ApJ...946...43W}.

First, we calculated the reddening curves characterized by the CER $E(\lambda-{\rm J400})/E({\rm J067}-{\rm J080})$ in a way similar to that described in Section~\ref{reddening}, 
where J400, J067, and J080 correspond to wavelengths of $4.0\mum$, $0.67\mum$, and $0.80\mum$, respectively. 
The choice of $4.0\mum$ as the reference wavelength is because the extinction at $4.0\mum$ is much lower than that at short wavelength. 
Therefore, the CER $E(\lambda-{\rm J400})/E({\rm J067}-{\rm J080})$ can be approximated to the relative extinction $A_\lambda/E({\rm J067}-{\rm J080})$ that is directly related to $\Rv$, resulting in more accurate $\Rv$ values. 
This method has also been used by \citet{2016ApJ...821...78S, 2019ApJ...886..108F, 2023ApJ...956...26L}.

In matching our reddening curves to $\Rv$ curves, we considered a set of $\Rv$-parameterized curves with $\Rv$ from 2.1 to 5.5 in a step of 0.01. 
The optimal $\Rv$ was determined by minimizing the RMSE. 
Half of the 384 pairs have low optical fluxes and large errors, resulting in a poor match. 
We excluded these pairs by the criterion of $2.1<\Rv<5.5$, remaining 188 pairs. 
All 188 pairs are in the Westerlund 2 region.
The low-extinction sources in the extinction pairs come from two sightlines with (R.A. decl.)=($120^{\circ}.04, -10^{\circ}.78$) and  (R.A. decl.)=($80^{\circ}.49, -69^{\circ}.50$). 
According to the 3D extinction map of \citet{2021ApJ...906...47G}, the maximum extinction $\Av$ towards these two sightlines are very small, only 0.15 mag and 0.36 mag. Due to the low extinction, the extinction curves of these two sightlines satisfying $\Rv=3.1$ or $\Rv=4.0$ have no effect on our results.
Finally, we obtained a weighted average $\Rv$ value of $3.93\pm0.23$. 
This value is consistent with the result of $\Rv=3.95\pm0.14$ of \citet{2015AJ....150...78Z} based on the Hubble Space Telescope multiband photometry.
Our $\Rv$ error is slightly large because of the low flux of the sample in the optical bands.
We also calculated the $\Rv$ using a single CER, i.e., $E({\rm J080}-{\rm J400})/E({\rm J067}-{\rm J080})$, and obtained a result of $\Rv=4.15\pm0.29$ consistent with the $\Rv$ obtained from the reddening curve.

Furthermore, we discuss the $\Rv$ values inferred from the different extinction curves. 
The average $\Rv$ values derived using the extinction curves of \citet{2023ApJ...946...43W}, \citet{1989ApJ...345..245C}, \citet{2019ApJ...886..108F}, and \citet{2023ApJ...950...86G} are $3.93\pm0.23$, $4.07\pm0.28$, $3.21\pm0.25$, and $3.54\pm0.22$, respectively. 
The $\Rv$ values based on the extinction laws of \citet{2023ApJ...946...43W} and \citet{1989ApJ...345..245C} are more consistent because both extinction laws are parameterized seven-order polynomial curves, and thus the $\Rv$ values are more smooth and less affected by the selection of the reference band. 
In contrast, the extinction laws of \citet{2019ApJ...886..108F} are non-parameterized, and the determination of $\Rv$ is highly dependent on the selection of reference bands. This is the reason that the $\Rv$ calculated based on \citet{2019ApJ...886..108F} is small. 
The extinction curves of \citet{2023ApJ...950...86G} are parameterized but has 47 parameters, resulting in local fluctuations. It yields larger $\Rv$ values than those of \citet{2019ApJ...886..108F} and closer to those of \citet{2023ApJ...946...43W} and \citet{1989ApJ...345..245C}. The $\Rv$ calculated by matching the CER with the four extinction laws would be more consistent if there is a reference wavelength shorter than 600 nm. However, it is not available for JWST spectra.

The right panel of Figure~\ref{fig:alpha_lambda} shows the distribution of $\Rv$ with $\alpha$. 
The distribution of the derived $\alpha$ satisfies a Gaussian distribution with a mean value of $\sim1.98$. 
We tested the correlation between $\alpha$ and $\Rv$. The correlation is very weak with the correlation coefficient $\rho=-0.0358$ and the associated probability $p=0.626$. The null hypothesis of significant correlation was rejected. It implies that $\alpha$ is an independent parameter that may need to be taken into account in future extinction law studies. For IR instruments such as JWST, $\alpha$ is a more appropriate parameter to indicate the extinction law. Based on $\alpha$, we then convert reddening curves into extinction curves.

Several studies have also discussed the correlation between $\alpha$ and $\Rv$. 
For example, \citet{2009ApJ...699.1209F} reported an anti-correlation between $\alpha$ and $\Rv$ based on extinction curves of 14 sightlines. 
Recently, \citet{2022ApJ...930...15D} studied extinction curves of 15 sightlines with $\AV<4$ mag and also found an anti-correlation between $\alpha$ and $\Rv$. 
However, \citet{2023arXiv231211762B} reanalyzed the correlation by using a mock analysis and argued that there is no correlation. 
They concluded the apparent correlation between $\alpha$ and $\Rv$ can be fully explained as arising from covariant uncertainties of these two parameters.

\subsection{The Near-IR Extinction Law}\label{Extcurve}

\begin{figure*}[ht]
\centering
\includegraphics[angle=0,width=6.5in]{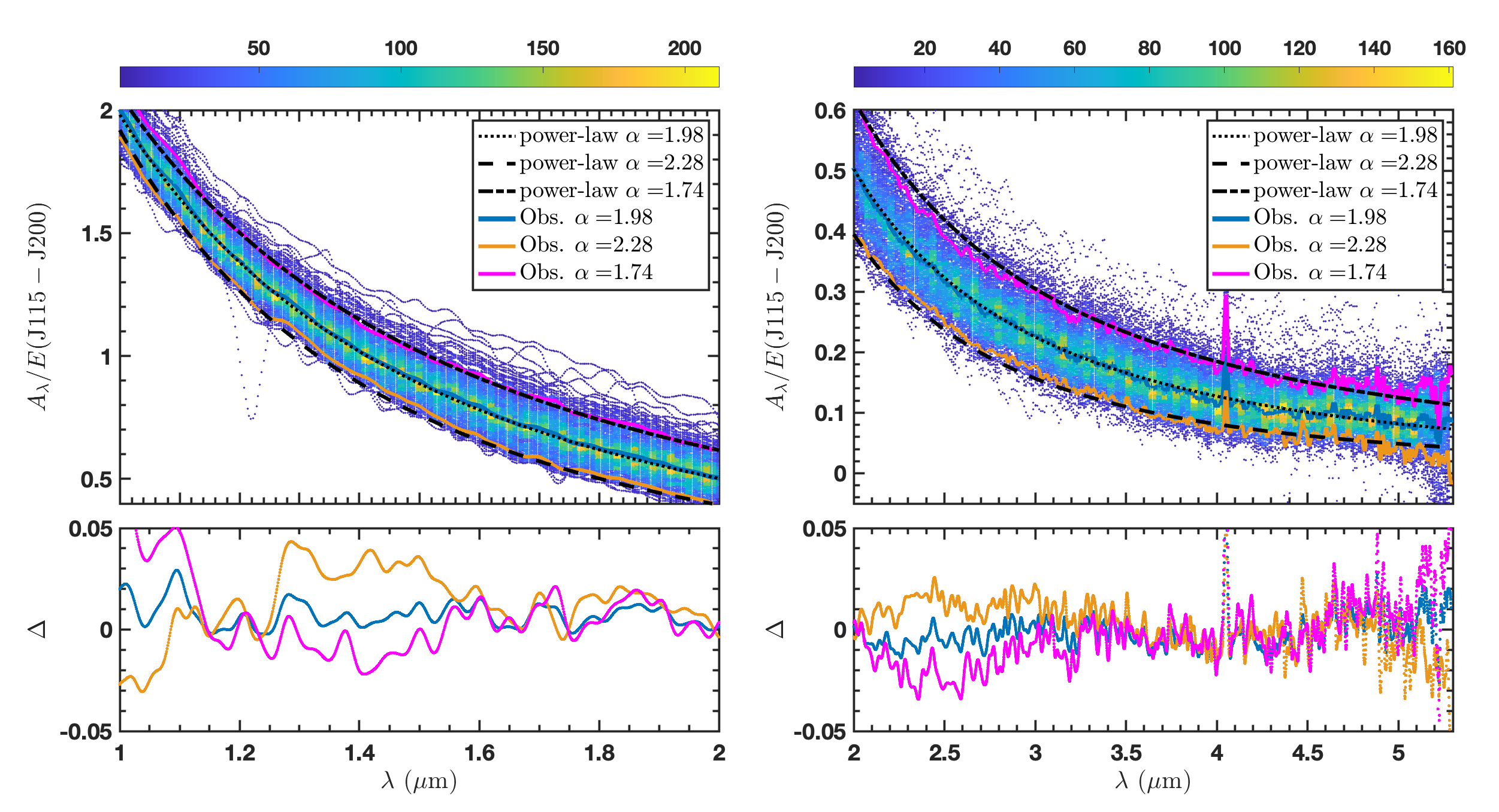}
\caption{
               \label{fig:ext}
Top: the determined extinction curves of 384 extinction pairs colored by the number density. 
The left and right panels are extinction curves of $1-2\mum$ and $2-5.3\mum$, respectively. 
The black lines are the power-law type curves with $\alpha$= 1.98 (dotted), 2.28 (long dashed), and 1.74 (dash-dotted), respectively. 
The bold blue, orange, magenta lines are the observed extinction curves for $\alpha$=1.98, 2.28, and 1.74, respectively. 
Bottom: the distribution of the difference $\Delta$ between the parameterized power-law type curves and the observed extinction curves. The blue, orange, magenta lines correspond to $\alpha$=1.98, 2.28, and 1.74, respectively.
} 
\end{figure*}

With the derived $\alpha$ value and Equation~\ref{equ_A/E}, we covert the reddening curves into the extinction curves. 
The top panel of Figure~\ref{fig:ext} displays the determined extinction curves of 384 extinction pairs colored by the number density. 
To better illustrate the details, the left and right panels of Figure~\ref{fig:ext} show the extinction curves of $1-2\mum$ and $2-5.3\mum$, respectively.
The black lines are the power-law type curves with given $\alpha$ values. 
The dotted, long dashed, and dash-dotted lines correspond to $\alpha$ equal to 1.98, 2.28, and 1.74, respectively. 
The bold-colored lines are the observed extinction curves for different $\alpha$ values. 
The bold blue line is the observed extinction curve with $\alpha=1.98$, which is the average extinction curve of 384 extinction pairs. 
The bold orange line is the observed extinction curve with $\alpha=2.28$, which is the average extinction curve of 43 pairs with $2.20<\alpha<2.39$. It is close to the lower edge of the curves.
The bold magenta line is the observed extinction curve with $\alpha=1.74$, which is the average extinction curve of 28 pairs with $1.64<\alpha<1.80$. It is close to the upper edge of all curves.
As shown in the top panel of Figure~\ref{fig:ext}, the distribution of the derived extinction curves is broad. Such a wide distribution not only includes curves with $\alpha$ from $\sim1.6$ to 2.4, but also could include curves that do not satisfy the power-law function. This also demonstrates that our method does not rely on a priori extinction law.

The bottom panel of Figure~\ref{fig:ext} displays the distribution of the difference $\Delta$ between the theoretical laws and observed curves. Different colored lines correspond to results with different $\alpha$. Overall, the observed extinction curves and theoretical extinction laws are in good agreement with a dispersion of less than 0.05. 
This means that when correcting the near-IR extinction, it is appropriate to use a simplified power-law type extinction law.
At some wavelengths, such as $\sim1.4\mum$ and $\sim2.5\mum$, the dispersion is slightly larger than others. Further corrections to them require more complex extinction curves (see Section \ref{alpha_extcurve}).

\section{Discussion}\label{discussion}
\subsection{The Comparison of the Power-Law Index $\alpha$}\label{comparison}

The near-IR $0.9-3\mum$ extinction curve can be expressed as a power-law $A_\lambda \propto \lambda^{-\alpha}$, as described in Section~\ref{intro}. 
Over the last four decades, the reported index ranges from 1.6 to 2.6. Whether the index varies with line of sight, the optical extinction $\Av$, and the choice of wavelength region is controversial. 
In Section~\ref{alpha}, we used a power-law to fit 408 reddening curves in the IR range of $1-5.3\mum$ to obtain the index $\alpha$. 
Among them, 384 curves were well fitted with a fitting error of $\sim 0.03$. 
This suggests that the near-IR $1-5.3\mum$ extinction can be well represented by a power-law $A_\lambda \propto \lambda^{-\alpha}$. 
 
As shown in the top panel of Figure~\ref{fig:ext}, our derived index $\alpha$ ranges from 1.51 to 2.39, with a mean value of 1.98. 
This value is close to the average value of the Galactic diffuse regions, e.g., 1.95 \citep{2014ApJ...788L..12W}, 2.07 \citep{2019ApJ...877..116W}. 
However, it is higher than the average value of 1.71 for the 15 OB star sightlines recently measured by \citet{2022ApJ...930...15D} using IRTF/SpeX near-IR ($0.8-5.5\mum$) spectra.
\citet{2022ApJ...930...15D} obtained the index in the ranges of $1.36-2.2$, with significant variations from sightline to sightline.

The dispersion of 384 $\alpha$ values is 0.15 larger than the mean error of 0.08 for each pair. 
It implies that the near-IR extinction curve may be variable, but it is also possible that the assumption of a power-law type extinction law is too simple, which leads to an underestimation of the error.  
More JWST spectra will be needed in the future to determine whether the near-IR extinction law is variable. 
We discuss the variation of extinction curves and establish the non-parameterized $\alpha$-dependent extinction curves in the next section.

\subsection{The non-parameterized $\alpha$-Dependent Extinction Curves}\label{alpha_extcurve}
In this section, we assume that the extinction curves vary with $\alpha$ and try to establish the non-parameterized $\alpha$-dependent extinction curves in the wavelength range of $0.6-5.3\mum$. 
The idea is that we obtain the average $\alpha$ extinction curve and then analyze the change in the extinction curve at each wavelength due to the change in $\alpha$. This is similar to that used by \citet{2019ApJ...886..108F} to derive the $\Rv$-dependent extinction curves. 
To characterize how the CER $k_\lambda$ varies with $\alpha$ at different wavelength $\lambda$, 
we fit $k_\lambda$ and $\alpha$ with a linear function  
\begin{eqnarray} \label{equ_slope}
k_\lambda  = a_\lambda*\alpha+b_\lambda~,  
\end{eqnarray}
where $a_\lambda$ and $b_\lambda$ are the slope and intercept of the fit, respectively. 
The slope $a_\lambda$ represents the variation of $k_\lambda$ with $\alpha$ at a given wavelength.

\begin{figure*}[ht]
\centering
\includegraphics[angle=0,width=6.0in]{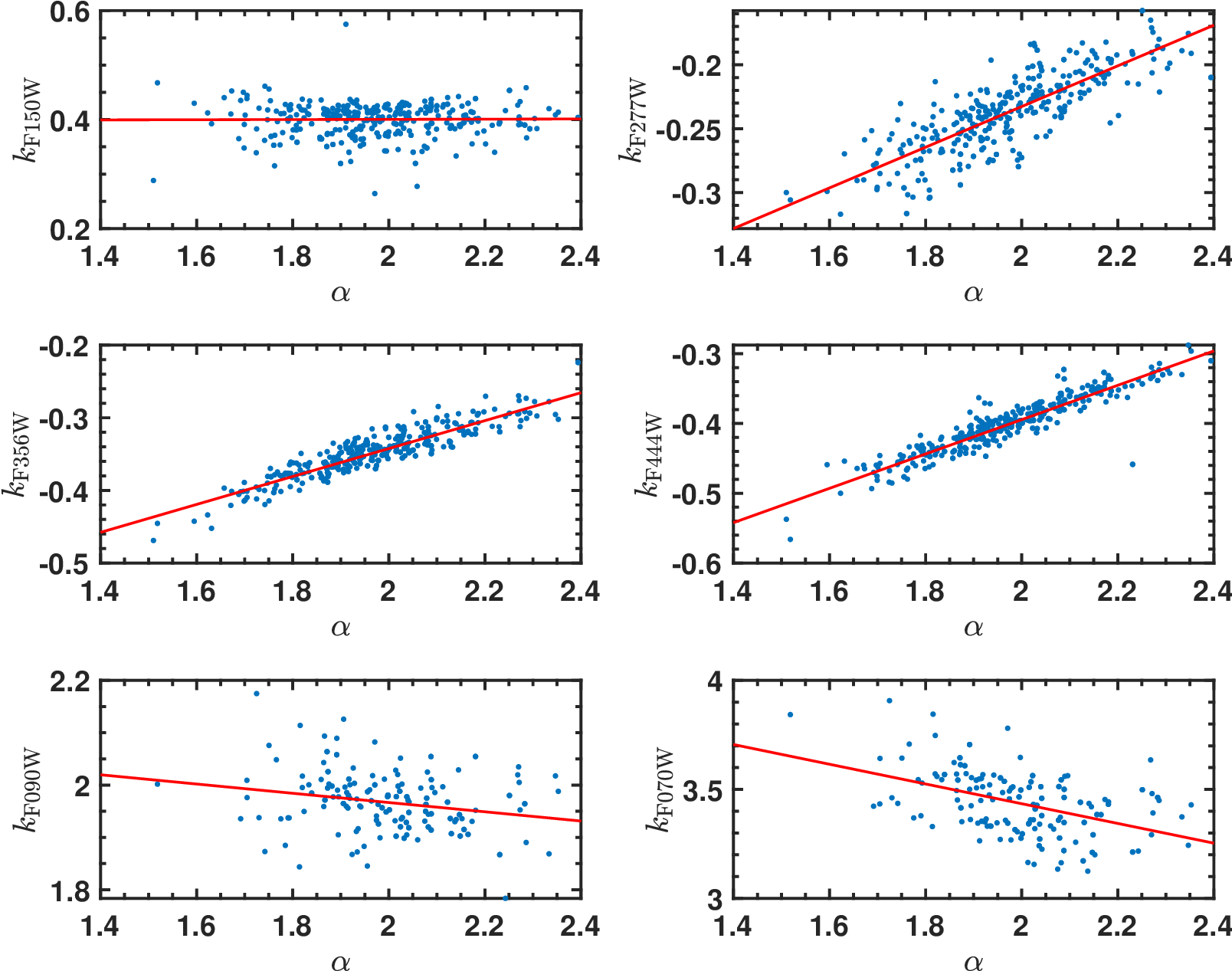}
\caption{
               \label{fig:Krelation}
The distribution of the CER $k_\lambda$ with the near-IR power-law index $\alpha$ at six JWST NIRcam bands, where $\lambda$ are F150W, F277W, F356W, F444W, F090W, and F070W, respectively, from the top left to the bottom right.
The red lines are the linear fit lines.} 
\end{figure*}

We first investigated the variation of CER $k_\lambda$ with $\alpha$ at six broadbands of JWST NIRcam, including F070W, F090W, F150W, F277W, F356W, and F444W.  
For each JWST band, we determined the value of $k_\lambda$ at the effective wavelength $\lambda_{\rm eff}$. 
Figure~\ref{fig:Krelation} shows the distribution of  $k_\lambda$ with $\alpha$ at each band. 
In the IR bands, $k_\lambda$ is significantly correlated with $\alpha$.  
At wavelength $>2\mum$, the larger the $\alpha$ is, the larger the $k_\lambda$ is.  
In the optical bands with wavelengths $< 1\mum$, such as F070W and F090W, $k_\lambda$ is also correlated with $\alpha$, but the scatter is larger. These demonstrate that $\alpha$ is an appropriate parameter to establish JWST bands' extinction laws.

\begin{figure*}[ht]
\centering
\includegraphics[angle=0,width=6.5in]{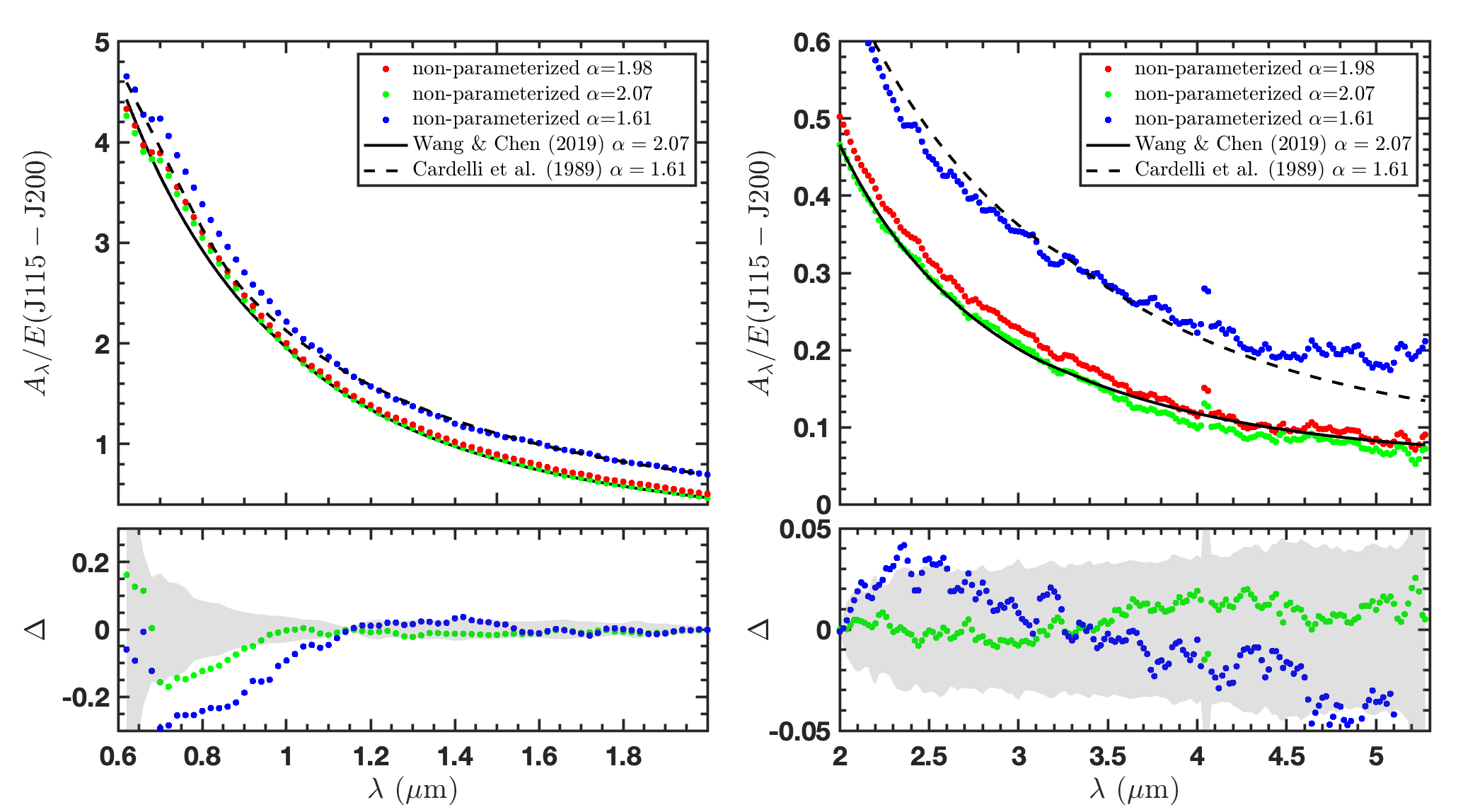}
\caption{
               \label{fig:alpha_ext}
Top: the extinction curves for $0.6-2.0\mum$ (left) and $2.0-5.3\mum$ (right), respectively.
Red, green, and blue dotted lines are our non-parameterized $\alpha$-dependent extinction curves with $\alpha=1.98$, 2.07, and 1.61, respectively. 
Black solid lines and black dashed lines are the parameterized extinction laws of \citet[][$\alpha=2.07$]{2019ApJ...877..116W} and \citet[][$\alpha=1.61$]{1989ApJ...345..245C}, respectively.  
Bottom: The differences $\Delta$ between the power-law extinction laws and our non-parameterized $\alpha$-dependent extinction curves. 
Green dotted lines and blue dotted lines are for $\alpha=2.07$ and $\alpha=1.61$, respectively.  
The grey shadow presents the $1\sigma$ scatter of the $\alpha=1.98$ extinction curve. } 
\end{figure*}

We further quantified the variation of $k_\lambda$ with $\alpha$ over the wavelength range of 0.6 to $5.3\mum$. We binned the wavelengths with a bin size of 10 nm, which yielded 468 bins between 0.6 and $5.3\mum$. 
After that, we calculated the average CER $<k_\lambda>$ within each bin. 
We fitted $<k_\lambda>$ versus $\alpha$ at each bin by using Equation~\ref{equ_slope} and recorded the corresponding slopes and slope errors. 
Finally, for a given $\alpha$, we obtained the corresponding $<k_\lambda'>$ reddening curve and $\frac{A_\lambda}{E(\rm J115-J200)}$ extinction curve. 
The red, green, and blue dotted lines in the top panel of Figure~\ref{fig:alpha_ext} are the derived non-parameterized $\alpha$-dependent extinction curves with $\alpha=1.98$, $\alpha=2.07$, and $\alpha=1.61$, respectively. 
The value of $\alpha=1.98$ is the average value that we obtained in Section~\ref{Extcurve}. 
The values of $\alpha=2.07$ and $\alpha=1.61$ are from the Galactic average extinction laws of \citet{2019ApJ...877..116W} and \citet{1989ApJ...345..245C}, respectively. As $\alpha$ decrease, the extinction curves become flatter, i.e. the relative extinction at longer wavelengths increases.
For comparison, we also plot the parameterized extinction laws of \citet[][black solid line, $\alpha=2.07$]{2019ApJ...877..116W} and \citet[][black dashed line, $\alpha=1.61$]{1989ApJ...345..245C}. 

The differences $\Delta$ between the parameterized power-law type extinction laws and our non-parameterized extinction curves are plotted in the bottom panel of Figure~\ref{fig:alpha_ext} as the green dotted lines for $\alpha=2.07$ and blue dotted lines for $\alpha=1.61$.  
The grey shadow presents the $1\sigma$ scatter of the $\alpha=1.98$ extinction curve. 

At wavelengths longer than $1\mum$, our non-parameterized extinction curves are consistent with the previous parameterized extinction curves at the same $\alpha$. The differences in $\frac{A_\lambda}{E(\rm J115-J200)}$ are less than 0.02 and 0.05, respectively, compared to the parameterized near-IR extinction curves of $\alpha=2.07$ and $\alpha=1.61$. At wavelengths shorter than $1\mum$, the differences are larger, but percentage difference is still less than 5\%.

We smooth the curves and provide non-parameterized $\alpha$-dependent extinction curves of $\frac{A_\lambda}{E(\rm J115-J200)}$ with $\alpha$ ranging from 1.5 to 2.4 and wavelengths ranging from 0.6 to $5.3\mum$. Table~\ref{tab:ext_alpha} is an example. The columns are the power-law index $\alpha$, the wavelength $\lambda$, the extinction coefficients $\frac{A_\lambda}{E(\rm J115-J200)}$, and $\frac{A_\lambda}{A_{\rm J200}}$, respectively. 
For 384 extinction pairs, our established non-parameterized $\alpha$-dependent extinction curves have a mean standard deviation of 0.0267 compared to observations. We used $\sigma_\lambda/\sqrt{n}$ as the error in the mean extinction curve ($\alpha=1.98$) and calculated the error in the other $\alpha$ extinction curves by considering the error in the CER slope $k_\lambda$ (see Section~\ref{alpha}). The errors of the extinction coefficients $\frac{A_\lambda}{E(\rm J115-J200)}$ and $\frac{A_\lambda}{A_{\rm J200}}$ are also listed in Table~\ref{tab:ext_alpha}. 

The non-parameterized $\alpha$-dependent extinction curve has two advantages over the parameterized power-law type extinction law. 
One is that the non-parameterized $\alpha$-dependent extinction curves provide reliable extinction coefficients shorter than $1\mum$, which can be better applied to the extinction correction of JWST-based photometry and spectra. The second is that the non-parameterized $\alpha$-dependent extinction curves match the observations better. For 384 extinction pairs and wavelengths longer than $1\mum$, the mean value of the standard deviation of the differences is 0.019, which is slightly smaller than the value of the parameterized power-law type extinction law (0.020). If the median value of the standard deviation of the percentage difference is considered, it is 7\% and 8\% for the non-parameterized $\alpha$-dependent extinction curves and the parameterized power-law type extinction laws, respectively.

\subsection{The Extinction Coefficients in the JWST NIRcam Bands}\label{JWSText}

Based on our determined non-parameterized $\alpha$-dependent extinction curves, we estimate the extinction coefficients $\frac{A_\lambda}{A_{\rm F200W}}$ and $\frac{A_\lambda}{E(\rm F115W-F200W)}$ for the bandpasses of the JWST NIRcam. 
Table~\ref{tab:ext_JWST} displays the derived extinction coefficients at three typical $\alpha$ values, i.e., 1.61, 1.98, and 2.07. 
The adopted effective wavelength $\lambda_{\rm eff}$ and CERs $\frac{E(\lambda-{\rm F200W})}{E({\rm F115W-F200W})}$ are also listed in Table~\ref{tab:ext_JWST}.  

\subsection{The Priori Extinction Laws}\label{prioriextlaw}

We also tried to use different priori extinction laws to select extinction pairs. For example, when using the  extinction law of \citet[][$\alpha=1.61$]{1989ApJ...345..245C}, only 119 extinction pairs were obtained. 151 extinction pairs were obtained based on a priori $\alpha=2.50$ extinction law. When using the flat IR extinction law of \citet{2015ApJ...811...38W}, only 115 extinction pairs were obtained. In addition, using the flat IR extinction law as the priori, the determined extinction curves still satisfy the power-law type curves (see Appendix Figure ~\ref{fig:ext_alpha_WLG}, the thick black solid line is the adopted flat IR extinction law). Compared to $\alpha=2.07$ extinction law, the significantly reduced pairs imply that these extinction laws are not optimal.

To determine the a priori extinction law, we tried different power-law extinction law with $\alpha = 0.9-3.3$ as the priori and obtained all extinction pairs satisfy $|\Delta\Av|>6.2$ mag and RMSE $<0.5$, where $\alpha$ is in a step of 0.2. We obtained a total of 1,251 pairs without duplicates, and the average $\alpha$ calculated based on them is $1.99\pm0.46$, which is consistent with our $1.98\pm0.15$ determined in Section \ref{alpha}. Appendix Figure~\ref{fig:alpha_RMSE} show the distribution of RMSE with different priori $\alpha$. Based on this figure, we find that more optimal extinction pairs can be selected using the priori extinction law of $\alpha\sim2.1$. These indicate that taken the $\alpha=2.07$ extinction law of \citet{2019ApJ...877..116W} is a suitable priori in this work to study the Galactic extinction law based on JWST spectra. 

\section{Conclusion}\label{conclusion}

Based on the available 993 prism/CLEAR spectra (resolution $\sim$100) from the JWST NIRSpec data, we have investigated the optical to IR interstellar dust extinction law. 
We propose a pair method to obtain the 0.6 to $5.3\mum$ reddening curves based only on JWST observed spectra. 
With the derived near-IR power-law index $\alpha$, we converted the reddening curves into the extinction curves.
The main results of this work are as follows.
\begin{enumerate} 
\item
Considering the difficulty of obtaining atmospheric parameters and the uncertainties in the IR spectra of red stars, in this work, we tried to build extinction pairs based on observed JWST spectra with very small or even negligible extinction. 
Each extinction pair contains the extinction SED $F_\lambda^\star$ and the corresponding non-extinction SED $F_{\lambda}^0$. 
This pair method also has a low dependence on a priori extinction law and is very suitable for JWST observations.  
\item
We calculated the CER $\frac{E(\lambda- {\rm J200})}{E({\rm J115}-{\rm J200})}$ to obtain the reddening curves based on the monochromatic near-IR J115 ($1.15\mum$) and J200 ($2.0\mum$) bands. 
We found that the observed reddening curves in the IR $1.0-5.3\mum$ can be well described by a power-law $A_\lambda \propto \lambda^{-\alpha}$.
By fitting it with a power-law, we obtained the index $\alpha$. 
The resulting $\alpha$ value ranges from 1.51 to 2.39 with an average value of $\alpha=1.98\pm0.15$. 
We used the derived $\alpha$ value to convert the reddening curves into extinction curves. 
\item
We also estimated the $\Rv$ value by matching the reddening curves in the optical $0.61-1\mum$ with the $\Rv$-parameterized extinction curves of \citet{2023ApJ...946...43W}. 
The weighted average $\Rv$ value of $3.93\pm0.23$ is consistent with the previously reported value for the Westerlund 2 region. 
The correlation between $\Rv$ and $\alpha$ is very weak, which means $\alpha$ is most likely an independent parameter. 
For IR instruments such as JWST, it is more appropriate to establish an extinction curve based on $\alpha$ than on $\Rv$.  
\item
We discussed the variation of the extinction curves and established the non-parameterized $\alpha$-dependent extinction curve in the wavelength range of $0.6-5.3\mum$. 
The non-parameterized $\alpha$-dependent extinction curve describes the JWST 0.6 to $5.3\mum$ observed extinction curve more accurately than the parameterized power-law type extinction law in two aspects:   
1) At $\lambda < 1\mum$, the parameterized power-law type extinction law deviates significantly from the observations, while the non-parameterized $\alpha$-dependent extinction curve represents the observations well. 
The non-parameterized $\alpha$-dependent extinction curves have a mean standard deviation of 0.0267 compared to observations. 
2)  At IR wavelengths of $1 < \lambda < 5.3\mum$, both the non-parameterized $\alpha$-dependent extinction curve and the parameterized power-law type extinction law can describe the observed extinction curve, but the non-parameterized $\alpha$-dependent extinction curves are more consistent with the details of the observations.  
Therefore, the non-parameterized $\alpha$-dependent extinction curves can be better applied to the extinction correction of JWST-based photometry and spectra.
The extinction coefficients $\frac{A_\lambda}{E(\rm J115-J200)}$ and $\frac{A_\lambda}{A_{\rm J200}}$ for different $\alpha$ in the wavelength range of 0.6 to $5.3\mum$ are provided in Table~\ref{tab:ext_alpha}.  
\item 
Based on the derived non-parameterized $\alpha$-dependent extinction curves, we provided the extinction coefficients $\frac{A_\lambda}{A_{\rm F200W}}$ and $\frac{A_\lambda}{E(\rm F115W-F200W)}$ for the JWST NIRCam bandpasses (Table~\ref{tab:ext_JWST}). 
\end{enumerate}

\section*{Acknowledgements}     
We thank the referee for very insightful and helpful suggestions/comments. 
This work is supported by the National Natural Science Foundation of China (NSFC) through the projects 12373028, 12322306, 12173047, 12003046, and 12133002.  
This work is also supported by the science research grants from the China Manned Space Project with No. CMS-CSST-220221-A09 and the National Key Research and Development Program of China, grant 2019YFA0405504. 
S.W. and X.C. acknowledge support from the Youth Innovation Promotion Association of the CAS (grant No. 2023065 and 2022055). 

This work is based on observations made with the NASA/ ESA/CSA James Webb Space Telescope. The data were obtained from the Mikulski Archive for Space Telescopes (MAST) at the Space Telescope Science Institute. The specific observations analyzed can be accessed via \dataset[https://doi.org/10.17909/0f0s-0531]{https://doi.org/10.17909/0f0s-0531}.  

Facility: JWST (NIRSpec).

\bibliography{reference}{}
\bibliographystyle{aasjournal}

\begin{table*}[ht]
\begin{center}
\caption{\label{tab:ext_alpha} Non-parameterized $\alpha$-Dependent Extinction Coefficients in $0.6-5.3\mum$} 
\begin{tabular}{cccc}
\hline \hline                                              
$\alpha$  & $\lambda^*$ ($\mum$)   
& $\frac{A_\lambda}{E(\rm J115-J200)}$ &  $\frac{A_\lambda}{A_{\rm J200}}$  \\ 
\hline
1.50 &  0.60 &  $4.8647 \pm0.0873$  &  $6.2926 \pm0.1129$   \\  
1.50 &  0.61 &  $4.8024 \pm0.0737$  &  $6.2120 \pm0.0953$   \\  
1.50 &  0.62 &  $4.7412 \pm0.0515$  &  $6.1328 \pm0.0667$   \\  
1.50 &  0.63 &  $4.6807 \pm0.0467$  &  $6.0546 \pm0.0604$   \\  
1.50 &  0.64 &  $4.6205 \pm0.0445$  &  $5.9767 \pm0.0576$   \\  
1.50 &  0.65 &  $4.5602 \pm0.0416$  &  $5.8988 \pm0.0539$   \\  
1.50 &  0.66 &  $4.4995 \pm0.0414$  &  $5.8202 \pm0.0535$   \\  
1.50 &  0.67 &  $4.4380 \pm0.0331$  &  $5.7406 \pm0.0428$   \\  
1.50 &  0.68 &  $4.3760 \pm0.0251$  &  $5.6605 \pm0.0325$   \\  
1.50 &  0.69 &  $4.3137 \pm0.0247$  &  $5.5799 \pm0.0320$   \\  
...  &  ...  &  ...             &  ...        \\  
2.40 &  5.21 &  $0.0445 \pm0.0057$  &  $0.1233 \pm0.0158$   \\  
2.40 &  5.22 &  $0.0446 \pm0.0062$  &  $0.1238 \pm0.0172$   \\  
2.40 &  5.23 &  $0.0448 \pm0.0063$  &  $0.1242 \pm0.0176$   \\  
2.40 &  5.24 &  $0.0449 \pm0.0062$  &  $0.1246 \pm0.0172$   \\  
2.40 &  5.25 &  $0.0451 \pm0.0063$  &  $0.1251 \pm0.0174$   \\  
2.40 &  5.26 &  $0.0453 \pm0.0065$  &  $0.1256 \pm0.0179$   \\  
2.40 &  5.27 &  $0.0454 \pm0.0063$  &  $0.1260 \pm0.0175$   \\  
2.40 &  5.28 &  $0.0456 \pm0.0056$  &  $0.1265 \pm0.0155$   \\  
2.40 &  5.29 &  $0.0458 \pm0.0060$  &  $0.1270 \pm0.0165$   \\  
2.40 &  5.30 &  $0.0459 \pm0.0061$  &  $0.1274 \pm0.0169$   \\
\hline  
\end{tabular}
\tablenotetext{*}{J115 and J200 are the monochromatic near-IR bands at $1.15\mum$ and $2.0\mum$.} 
\tablenotetext{}{A complete table of extinction coefficients for different $\alpha$ is available online.}
\end{center}
\end{table*}

\begin{sidewaystable*}[ht]
\begin{center}
\caption{\label{tab:ext_JWST} Color Excess Ratios and Extinction Coefficients of JWST NIRcam Bands}\scriptsize 
\begin{tabular}{ccccccccccc}
\hline \hline                                              
Band  & $\lambda_{\rm eff}^*$    
& $\frac{E(\lambda-\rm F200W)}{E(\rm F115W-F200W)}$ &  $\frac{A_\lambda}{A_{\rm F200W}}$  &  $\frac{A_\lambda}{E(\rm F115W-F200W)}$  
& $\frac{E(\lambda-\rm F200W)}{E(\rm F115W-F200W)}$ &  $\frac{A_\lambda}{A_{\rm F200W}}$  &  $\frac{A_\lambda}{E(\rm F115W-F200W)}$  
& $\frac{E(\lambda-\rm F200W)}{E(\rm F115W-F200W)}$ &  $\frac{A_\lambda}{A_{\rm F200W}}$  &  $\frac{A_\lambda}{E(\rm F115W-F200W)}$\\ 
\cmidrule(r){1-2}  \cmidrule(lr){3-5}  \cmidrule(lr){6-8} \cmidrule(lr){9-11}
 &  $\mum$  & \multicolumn{3}{c}{$\alpha=1.61$}  &  \multicolumn{3}{c}{$\alpha=1.98$}  &  \multicolumn{3}{c}{$\alpha=2.07$} \\ 
\hline
F070W  &  0.70 &  $ 3.341 \pm0.019$  &  $5.786 \pm0.032$  &  $4.039\pm0.023$  &  $3.217 \pm0.010$  &  $7.321\pm0.022$  &  $3.726\pm0.011$  &  $3.185 \pm0.010$  &  $7.742\pm0.026$  &  $3.657\pm0.012$ \\          
F090W  &  0.90 &  $ 1.964 \pm0.007$  &  $3.814 \pm0.014$  &  $2.663\pm0.010$  &  $1.930 \pm0.003$  &  $4.791\pm0.007$  &  $2.439\pm0.004$  &  $1.920 \pm0.003$  &  $5.065\pm0.009$  &  $2.393\pm0.004$ \\          
F115W  &  1.14 &     1               &  $2.433 \pm0.001$  &  $1.698\pm0.001$  &    1               &  $2.965\pm0.001$  &  $1.509\pm0.000$  &    1               &  $3.117\pm0.001$  &  $1.472\pm0.000$ \\          
F150W  &  1.49 &  $ 0.379 \pm0.001$  &  $1.542 \pm0.005$  &  $1.077\pm0.004$  &  $0.381 \pm0.001$  &  $1.750\pm0.003$  &  $0.890\pm0.001$  &  $0.383 \pm0.001$  &  $1.810\pm0.003$  &  $0.855\pm0.002$ \\          
F200W  &  1.97 &     0               &    1               &  $0.698\pm0.001$  &    0               &    1              &  $0.509\pm0.000$  &    0               &    1              &  $0.472\pm0.001$ \\          
F277W  &  2.73 &  $ -0.283\pm0.001$  &  $0.595 \pm0.003$  &  $0.415\pm0.002$  &  $-0.242\pm0.001$  &  $0.524\pm0.002$  &  $0.267\pm0.001$  &  $-0.231\pm0.001$  &  $0.511\pm0.002$  &  $0.242\pm0.001$ \\          
F356W  &  3.53 &  $ -0.421\pm0.003$  &  $0.397 \pm0.003$  &  $0.277\pm0.002$  &  $-0.353\pm0.002$  &  $0.306\pm0.001$  &  $0.156\pm0.001$  &  $-0.335\pm0.002$  &  $0.291\pm0.002$  &  $0.137\pm0.001$ \\          
F444W  &  4.35 &  $ -0.499\pm0.004$  &  $0.286 \pm0.002$  &  $0.200\pm0.002$  &  $-0.411\pm0.003$  &  $0.193\pm0.001$  &  $0.098\pm0.001$  &  $-0.388\pm0.004$  &  $0.179\pm0.002$  &  $0.084\pm0.001$ \\          
F150W2 &  1.48 &  $ 0.388 \pm0.001$  &  $1.556 \pm0.005$  &  $1.086\pm0.004$  &  $0.393 \pm0.001$  &  $1.772\pm0.003$  &  $0.902\pm0.001$  &  $0.395 \pm0.001$  &  $1.836\pm0.003$  &  $0.867\pm0.002$ \\          
F322W2 &  3.07 &  $ -0.349\pm0.002$  &  $0.500 \pm0.003$  &  $0.349\pm0.002$  &  $-0.294\pm0.001$  &  $0.422\pm0.001$  &  $0.215\pm0.001$  &  $-0.279\pm0.001$  &  $0.408\pm0.002$  &  $0.193\pm0.001$ \\
F140M  &  1.40 &  $ 0.486 \pm0.002$  &  $1.696 \pm0.006$  &  $1.184\pm0.004$  &  $0.495 \pm0.001$  &  $1.972\pm0.003$  &  $1.004\pm0.002$  &  $0.498 \pm0.001$  &  $2.054\pm0.004$  &  $0.970\pm0.002$ \\          
F162M  &  1.62 &  $ 0.262 \pm0.001$  &  $1.375 \pm0.005$  &  $0.960\pm0.003$  &  $0.244 \pm0.000$  &  $1.480\pm0.003$  &  $0.753\pm0.001$  &  $0.241 \pm0.000$  &  $1.510\pm0.003$  &  $0.713\pm0.001$ \\
F182M  &  1.84 &  $ 0.085 \pm0.000$  &  $1.121 \pm0.003$  &  $0.783\pm0.002$  &  $0.079 \pm0.000$  &  $1.154\pm0.002$  &  $0.588\pm0.001$  &  $0.078 \pm0.000$  &  $1.165\pm0.003$  &  $0.551\pm0.001$ \\
F210M  &  2.09 &  $ -0.071\pm0.000$  &  $0.898 \pm0.004$  &  $0.627\pm0.003$  &  $-0.064\pm0.000$  &  $0.874\pm0.002$  &  $0.445\pm0.001$  &  $-0.061\pm0.000$  &  $0.871\pm0.003$  &  $0.411\pm0.001$ \\
F250M  &  2.50 &  $ -0.222\pm0.001$  &  $0.682 \pm0.004$  &  $0.476\pm0.003$  &  $-0.192\pm0.001$  &  $0.623\pm0.002$  &  $0.317\pm0.001$  &  $-0.183\pm0.001$  &  $0.612\pm0.002$  &  $0.289\pm0.001$ \\
F300M  &  2.98 &  $ -0.332\pm0.002$  &  $0.525 \pm0.003$  &  $0.366\pm0.002$  &  $-0.280\pm0.001$  &  $0.450\pm0.002$  &  $0.229\pm0.001$  &  $-0.266\pm0.001$  &  $0.437\pm0.002$  &  $0.206\pm0.001$ \\
F335M  &  3.35 &  $ -0.397\pm0.002$  &  $0.431 \pm0.002$  &  $0.301\pm0.002$  &  $-0.331\pm0.001$  &  $0.349\pm0.001$  &  $0.178\pm0.001$  &  $-0.314\pm0.001$  &  $0.335\pm0.002$  &  $0.158\pm0.001$ \\
F360M  &  3.61 &  $ -0.432\pm0.003$  &  $0.382 \pm0.003$  &  $0.267\pm0.002$  &  $-0.362\pm0.002$  &  $0.288\pm0.001$  &  $0.147\pm0.001$  &  $-0.344\pm0.002$  &  $0.271\pm0.002$  &  $0.128\pm0.001$ \\
F410M  &  4.07 &  $ -0.473\pm0.009$  &  $0.322 \pm0.006$  &  $0.225\pm0.004$  &  $-0.394\pm0.003$  &  $0.225\pm0.002$  &  $0.115\pm0.001$  &  $-0.374\pm0.005$  &  $0.209\pm0.003$  &  $0.099\pm0.001$ \\
F430M  &  4.28 &  $ -0.493\pm0.005$  &  $0.293 \pm0.003$  &  $0.205\pm0.002$  &  $-0.408\pm0.003$  &  $0.199\pm0.002$  &  $0.101\pm0.001$  &  $-0.385\pm0.004$  &  $0.184\pm0.002$  &  $0.087\pm0.001$ \\
F460M  &  4.63 &  $ -0.510\pm0.007$  &  $0.269 \pm0.004$  &  $0.188\pm0.003$  &  $-0.417\pm0.004$  &  $0.180\pm0.002$  &  $0.092\pm0.001$  &  $-0.393\pm0.005$  &  $0.168\pm0.002$  &  $0.080\pm0.001$ \\
F480M  &  4.81 &  $ -0.522\pm0.005$  &  $0.253 \pm0.002$  &  $0.177\pm0.002$  &  $-0.425\pm0.003$  &  $0.166\pm0.001$  &  $0.085\pm0.001$  &  $-0.399\pm0.004$  &  $0.154\pm0.002$  &  $0.073\pm0.001$ \\
F164N  &  1.64 &  $ 0.243 \pm0.001$  &  $1.348 \pm0.005$  &  $0.941\pm0.004$  &  $0.225 \pm0.000$  &  $1.443\pm0.003$  &  $0.735\pm0.001$  &  $0.222 \pm0.001$  &  $1.470\pm0.003$  &  $0.694\pm0.002$ \\
F187N  &  1.87 &  $ 0.067 \pm0.000$  &  $1.096 \pm0.003$  &  $0.765\pm0.002$  &  $0.061 \pm0.000$  &  $1.120\pm0.002$  &  $0.570\pm0.001$  &  $0.061 \pm0.000$  &  $1.128\pm0.002$  &  $0.533\pm0.001$ \\
F212N  &  2.12 &  $ -0.087\pm0.000$  &  $0.876 \pm0.003$  &  $0.611\pm0.002$  &  $-0.078\pm0.000$  &  $0.847\pm0.002$  &  $0.431\pm0.001$  &  $-0.074\pm0.000$  &  $0.843\pm0.002$  &  $0.398\pm0.001$ \\
F323N  &  3.24 &  $ -0.396\pm0.002$  &  $0.433 \pm0.002$  &  $0.302\pm0.002$  &  $-0.319\pm0.001$  &  $0.373\pm0.001$  &  $0.190\pm0.001$  &  $-0.302\pm0.001$  &  $0.360\pm0.002$  &  $0.170\pm0.001$ \\
F405N  &  4.05 &  $ -0.471\pm0.018$  &  $0.325 \pm0.013$  &  $0.227\pm0.009$  &  $-0.393\pm0.016$  &  $0.228\pm0.009$  &  $0.116\pm0.005$  &  $-0.373\pm0.019$  &  $0.211\pm0.011$  &  $0.100\pm0.005$ \\
F466N  &  4.65 &  $ -0.511\pm0.006$  &  $0.267 \pm0.003$  &  $0.187\pm0.002$  &  $-0.418\pm0.004$  &  $0.179\pm0.002$  &  $0.091\pm0.001$  &  $-0.394\pm0.005$  &  $0.167\pm0.002$  &  $0.079\pm0.001$ \\
F470N  &  4.71 &  $ -0.515\pm0.006$  &  $0.263 \pm0.003$  &  $0.183\pm0.002$  &  $-0.420\pm0.005$  &  $0.175\pm0.002$  &  $0.089\pm0.001$  &  $-0.395\pm0.006$  &  $0.163\pm0.002$  &  $0.077\pm0.001$ \\
\hline  
\end{tabular}
\end{center}\normalsize
\tablenotetext{*}{$\lambda_{\rm eff}$ is the effective wavelength of JWST NIRcam bands based on Vega and the value is from the SVO Filter Profile Service, http://svo2.cab.inta-csic.es/theory/fps3/}. 
\end{sidewaystable*}

\clearpage

\appendix

\setcounter{table}{0}
\renewcommand{\thetable}{A\arabic{table}}
\setcounter{figure}{0}
\renewcommand{\thefigure}{A\arabic{figure}}

\section{Supplementary Figures}

\begin{figure*}[ht]
\centering
\includegraphics[angle=0,width=6.5in]{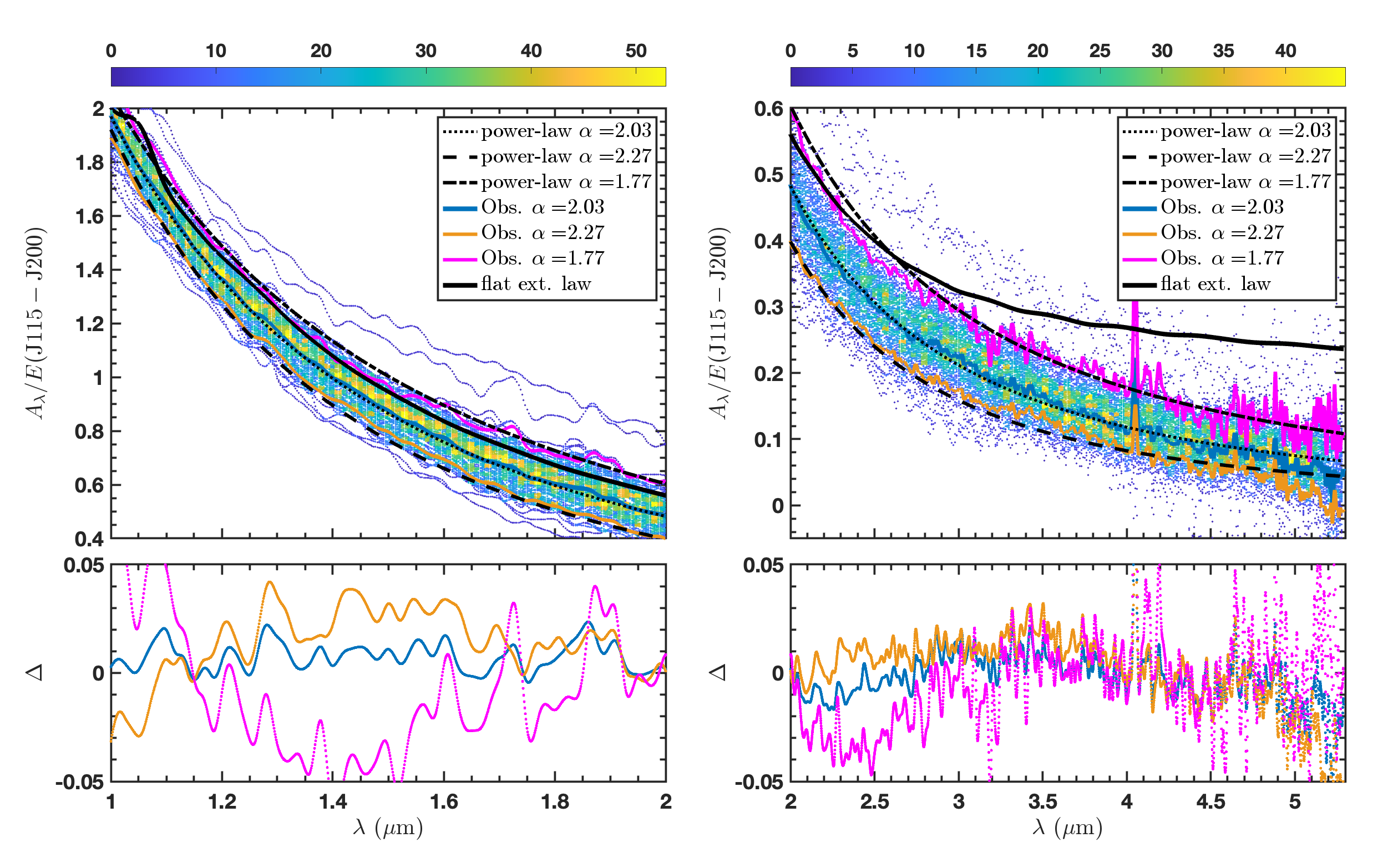}
\caption{
               \label{fig:ext_alpha_WLG}
The same as Figure~\ref{fig:ext}, but adopted the flat IR extinction law of \citet[][thick black solid line]{2015ApJ...811...38W} to find extinction pairs. 
Top: the determined extinction curves of the 115 selected extinction pairs colored by the number density. 
The left and right panels are extinction curves of $1-2\mum$ and $2-5.3\mum$, respectively. 
The black lines are the power-law type curves with $\alpha$= 2.03 (dotted), 2.27 (long dashed), and 1.77 (dash-dotted), respectively. 
The bold blue, orange, magenta lines are the observed extinction curves for $\alpha$=2.03, 2.27, and 1.77, respectively. 
Bottom: the distribution of the difference $\Delta$ between the parameterized power-law type curves and the observed extinction curves. The blue, orange, magenta lines correspond to $\alpha$=2.03, 2.27, and 1.77, respectively.}
 
\end{figure*}

\begin{figure*}[ht]
\centering
\includegraphics[angle=0,width=4.5in]{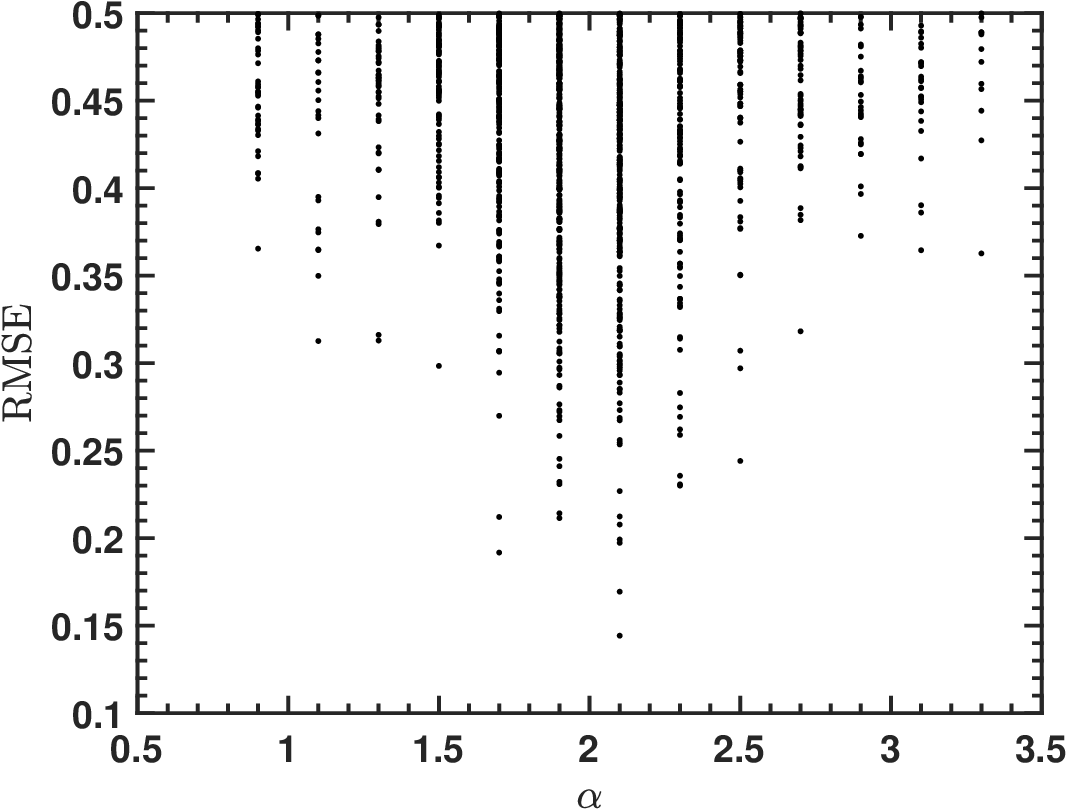}
\caption{\label{fig:alpha_RMSE}
The distribution of RMSE with different priori $\alpha$ for 1,251 extinction pairs, where RMSE is the minimum root mean square error of each extinction pair.
}
 
\end{figure*}

\end{document}